\documentclass[preprint,12pt]{elsarticle}

\usepackage{amssymb}
\pdfoutput=1

\usepackage{graphicx}
\usepackage{dcolumn}
\usepackage{bm}

\usepackage{hyperref}
\hypersetup{hypertex=true,
	colorlinks=true,
	linkcolor=blue,
	anchorcolor=blue,
	citecolor=blue}
\usepackage{color}
\usepackage{float}
\usepackage{siunitx}
\usepackage[dvipsnames]{xcolor}
\usepackage[defaultcolor=Maroon]{changes}
\definechangesauthor[name={Tian-Xue Ma},color=BlueGreen]{MTX}
\makeatletter
\def\ps@pprintTitle{%
	\let\@oddfoot\@empty
	\let\@evenfoot\@empty
}
\makeatother

\begin{document}

\begin{frontmatter}

\title{Topological rainbow trapping and broadband piezoelectric energy harvesting of acoustic waves in gradient phononic crystals with coupled interfaces}

\author[label1,label3]{Xiao-Lei Tang}
\author[label2]{Xue-Qian Zhang}
\author[label2]{Tian-Xue Ma\corref{cor1}}
\ead{matx@bjtu.edu.cn}
\author[label3]{Miso Kim}
\author[label1,label2]{Yue-Sheng Wang\corref{cor1}}
\ead{yswang@tju.edu.cn}
\cortext[cor1]{Corresponding authors}
\address[label1]{Department of Mechanics, School of Mechanical Engineering, Tianjin University, Tianjin 300350, PR China}
\address[label2]{Department of Mechanics, School of Physical Science and Engineering, Beijing Jiaotong University, Beijing 100044, PR China}
\address[label3]{School of Advanced Materials Science and Engineering, Sungkyunkwan University (SKKU), Suwon 16419, Republic of Korea}

\date{\today}	

\begin{abstract}
Topological phononic crystals (PCs) offer an innovative method for manipulating acoustic or elastic waves. In this study, we introduce the gradient PC structures with coupled interfaces, specifically designed to achieve topological rainbow trapping and broadband acoustic energy harvesting. By leveraging the geometric symmetry of PC unit cells, we merge two PCs with distinct topological phases to create coupled topological interfaces. Gradient modulation of structural parameters along the coupled interfaces induces rainbow trapping, where acoustic waves are spatially separated by frequency. The numerical and experimental results indicate that the acoustic waves of various frequencies are halted and magnified at distinct locations within the coupled interfaces. Compared to the bare harvester, the topological PC energy harvester markedly increases output power across a range of excitation frequencies, with a maximum amplification ratio of 91 observed in experiments. Furthermore, the topological rainbow trapping is robust against random structural disorders. The coupled interfaces exhibit broadband and multimodal capabilities, holding potential for various applications including selective filtering and enhanced sensing.

\end{abstract}


\end{frontmatter}


\section{\label{Sec1}Introduction}

\label{}

Utilizing acoustic energy to power small electronic devices is gaining popularity. Various transduction mechanisms, such as piezoelectric \cite{li2013low,choi2019brief,alqaleiby2024effects,eghbali2020enhancement,yuan2018low}, electromagnetic \cite{6731337,khan2016contributed,khan2018electromagnetic} and triboelectric \cite{yang2014triboelectrification,fan2015ultrathin,YU2022107205,SUN2023108430} approaches, have been utilized to transform acoustic energy into electrical energy. Though sounds are pervasive in daily life, their energy is typically characterized by low power density, limiting the efficiency of acoustic energy harvesting. To overcome this limitation, sounds need to be concentrated and localized at specified positions. Previous studies have reported several acoustic energy harvesters employing Helmholtz resonators \cite{peng2013enhanced} and 1/4-wavelength resonators \cite{li2013harvesting,ZHU2023108237}. Liu $et$ $al$. \cite{liu2008acoustic} achieved the acoustic energy harvesting by coupling a piezoelectric patch into the Helmholtz resonator. Yuan $et$ $al$. \cite{yuan20213d} proposed a 1/4-wavelength resonator-based acoustic triboelectric nanogenerator, which can produce 4.33 mW at 100 dB sound pressure.

Recently, with the studies of phononic crystals (PCs) in diverse disciplines, interests have also focused on their feasibility for wave energy harvesting. PCs are artificially designed materials with peculiar wave properties, including negative refraction \cite{zhu2014negative}, negative modulus \cite{liu2005analytic} and band gap \cite{goffaux2001theoretical}. Owing to their unique abilities to control acoustic or elastic waves, PCs have been extensively studied by the scientific community for sound transmission \cite{jiang2017transmission}, cloaking \cite{zheng2014acoustic} and energy harvesting \cite{ma2022energy,ma2020flexural,zhang2022topological,wang2018renewable}.  Incorporating a point defect into a pristine PC leads to the emergence of an acoustic localization effect due to the defect state of the PC. This means that acoustic waves gather at and near the point defect, resulting in a notable increase in pressure amplitude. The acoustic localization effect in PC structures can be harnessed for energy harvesting applications \cite{wu2009acoustic,gao2019acoustic,yang2015high}. Ma $et$ $al$. \cite{ma2021metamaterial} combined the PC and the Helmholtz resonator to improve the sound energy density by localizing and amplifying the waves. In addition, PC lenses \cite{jin2019gradient,lee2023acoustic} are a distinct type of gradient PC structures to regulate wave propagation, in which the effective refractive index of each unit cell is spatially distributed according to specific profiles.  Owing to their wave focusing ability, various PC lenses have been developed for the piezoelectric energy harvesting from elastic \cite{akbari2023defect,hyun2019gradient} or acoustic waves \cite{allam2021sound}. Kim $et$ $al$. \cite{kim2022gradient} integrated the Helmholtz resonance mechanism with a PC lens to achieve high power output for sounds below 1 kHz. Lee $et$ $al$. \cite{lee2022machine} designed an acoustic lens using machine learning to achieve sound wave energy harvesting.

The discovery of topological insulators has opened up unprecedented ways for the control of electromagnetic \cite{lu2014topological}, acoustic \cite{chen2023various,yin2022acoustic,zheng2022observation} and elastic \cite{ma2022topological,ma2022simple,ma2022flexural} waves. By arranging materials with diverse topological phases in accordance with bulk-edge correspondence, one can establish states that are topologically protected. Unlike traditional wave states, topological interface states (TISs) show remarkable durability in the face of structural defects. Even in the presence of defects or abrupt transitions, the TISs maintain smooth propagation across interfaces due to their inherent topological protection. Over the past decade, acoustic topological insulators have been utilized for unidirectional transmission \cite{xia2017topological,he2016acoustic}, logic operation\cite{lu2022multifunctional,pirie2022topological} and acoustic tweezers \cite{dai2021experimental,liu2022topological,du2023acoustic}. Besides, the rainbow trapping phenomenon of TISs has been reported, in which TISs at different frequencies are separated and halted at distinct locations. \cite{chaplain2020topological,tian2020dispersion,tang2023topological}. Moreover, due to their unique wave characteristics, topological PCs are a promising option for acoustic energy harvesting. Fan $et$ $al$. \cite{fan2019acoustic} proposed a one-dimensional (1D) PC tube with TISs to harvest the energy of acoustic waves. Zhao $et$ $al$. \cite{zhao2021subwavelength} utilized the presence of topological interface states within 1D Helmholtz resonator arrays for subwavelength acoustic energy harvesting.  Li  $et$ $al$. \cite{li2023acoustic} designed a topological acoustic energy harvester, where the TIS enhanced robustness and multi-resonant cavities were used to lower the operating frequency.  Also, several researchers have utilized TISs for energy harvesting of elastic waves \cite{lan2021energy,wen2022topological,chen2024rainbow,liu2024tunable,liu2023topologically}. Therefore, TISs exhibit excellent robustness and strong energy localization capabilities for acoustic/elastic waves amidst structural imperfections.

 However, the broadband energy harvesting of acoustic waves through topological coupled interfaces is rarely reported. Notably, in the regime of electromagnetic waves, Elshahat  $et$ $al$. \cite{elshahat2022broadband} observed the rainbow trapping of TISs in the coupled topological interfaces of gradient photonic crystal structures. Interestingly, the coupling of TISs has been found to exhibit broadband and multimodal characteristics. Here, inspired by the above-mentioned study, we present a graded PC structure incorporating coupled interfaces, specifically engineered for broadband topological rainbow trapping and efficient acoustic energy harvesting. The paper proceeds as follows. In Section \ref{Sec2}, the dispersion relations and topological properties of 2D PCs, as well as the group velocities of TISs, are investigated. Section \ref{Sec3} discusses the numerical and experimental findings on topological rainbow trapping and sound pressure amplification in designed PC structures with coupled interfaces. Section \ref{Sec4} reports the experimental validation of acoustic energy harvesting. The final conclusions are outlined in Section \ref{Sec5}.

\section{\label{Sec2}Band diagram analysis}

\begin{figure*}[]
    \centering
	\includegraphics{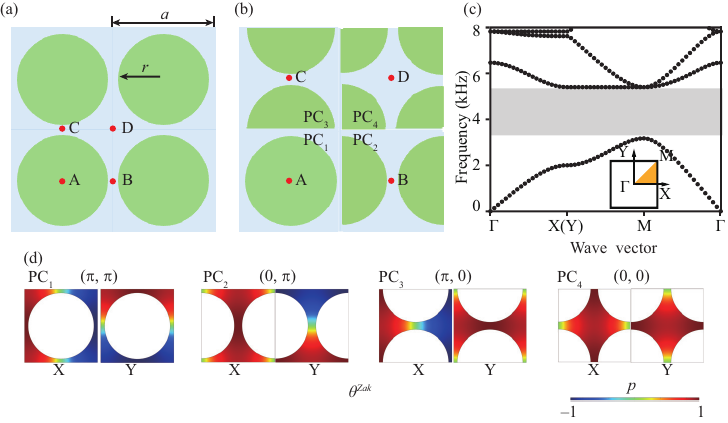}
	\caption{\label{Figure.1}(a) Depiction of the 2D square-lattice PC, in which polymer scatterers ( depicted as green areas) are arranged periodically against a light cyan background. (b) Four choices of the PC unit cells, with red points marking the various inversion centers. (c) Dispersion curves of the PC unit cells, with the band gap shaded in gray. (d) At the points X and Y, the pressure distributions of the first bulk band and the respective Zak phases ($\theta_x^{Zak}, \theta_y^{Zak}$) are presented.}
\end{figure*}

We design a 2D square-lattice PC, integrating polymeric scatterers (highlighted in green) within a background of air (depicted in light cyan), as sketched in Fig. \ref{Figure.1}(a). The PC structure is defined by the lattice constant $a$ and the radius $r$ of each circular scatterer. Given the pronounced acoustic impedance difference between the polymer materials and air, these polymeric scatterers act as rigid acoustic boundaries. The mass density and the sound speed of air are $\rho = 1.21$ $\rm{kg/m^3}$ and $c = 343$ m/s, respectively. Throughout this study, we employ the finite element method via COMSOL Multiphysics for simulations. Notably, the 2D PC has four inversion centers, labeled A, B, C, and D [see Fig. \ref{Figure.1}(a)]. According to the crystal periodicity, four types of PC unit cells can be obtained, as illustrated in Fig. \ref{Figure.1}(b). Fig. \ref{Figure.1}(c) shows the band diagram of the 2D PC, with a lattice constant of 45 mm and a scatterer radius of 0.45$a$. As depicted in Figs. \ref{Figure.1}(b) and \ref{Figure.1}(c), regardless of the four distinct unit cell selections, the band diagrams of the PC remain identical.

In this work, the 2D PC is protected by the inversion symmetry, restricting the Zak phase to either 0 or $\pi$. By studying the symmetry properties of the pressure field ($p$) distribution at the points (X and Y) in the Brillouin zone [see Fig. \ref{Figure.1}(d)], the corresponding ($\theta_x^{Zak}, \theta_y^{Zak}$) of four unit cells can be obtained \cite{xiao2015geometric}. Considering the mirror-symmetry of the pressure profile at point $\Gamma$, if the pressure profile at the points (i.e., X and Y) is symmetric$/$antisymmetric (S$/$A), the Zak phase $\theta_j^{Zak}$ of the first band is 0 (trivial phase)$/$$\pi$ (non-trivial phase) in the corresponding direction. Fig. \ref{Figure.1}(d) illustrates that the pressure fields ($p$) at X(Y) for the four unit cells are respectively A/A, S/A, A/S, and S/S, which leads to their associated 2D Zak phases ($\theta_x^{Zak}, \theta_y^{Zak}$) [($\pi$, $\pi$), (0, $\pi$), ($\pi$, 0) and (0, 0)]. In addition, the Zak phase of the first band of the 2D PC along the $j$ ($j = x, y$) direction is defined as $\theta_j^{Zak}$  =$ \int\rm$  $dk_xdk_y$ Tr$[A_j(k_x,k_y)]$, where $A_j(k_x,k_y) = \left\langle\psi\left|i\partial k_{j}\right|\psi\right\rangle$ is the Berry connection, $\psi$ denotes the periodic part of the Bloch pressure eigenfunction, with the integration performed over the first Brillouin zone \cite{asboth2016short,zangeneh2019topological,lu2021topological}. The numerically calculated Berry connections of different 2D PC unit cells are provided in the Supplementary Material. When two PCs with distinct topological phases (i.e., Zak phases) are placed adjacent to each other, the resulting PC structure generates topologically protected states at the interface (i.e., TISs). For an acoustic TIS, acoustic waves consistently consistently propagate along the interface of the PC structure and experience no backscattering.

\begin{figure*}[]
	\centering
	\includegraphics{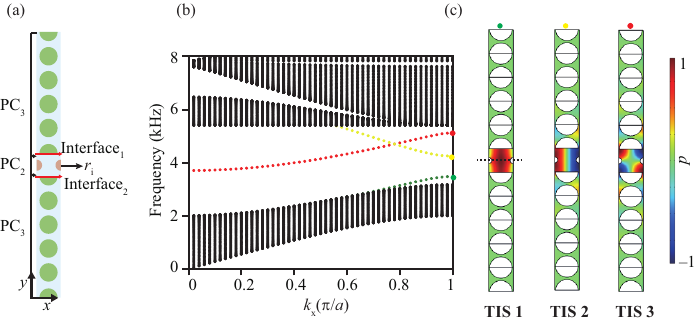}
	\caption{\label{Figure.2}(a) Depiction of the PC super-cell with two coupled interfaces, where a PC$_2$ unit is sandwiched between ten PC$_3$ units. (b) Band diagram of the PC super-cell, where three different TISs (marked by green, yellow and red dots, respectively) emerge inside the band gap. Bulk modes are indicated by black dots. (c) Pressure distributions of the three TISs at the Brillouin zone boundary $k_x = \pi/a$.}
\end{figure*}

Next, we integrate two varieties of PC unit cells with distinct topological phases ($\theta_x^{Zak} = 0, \theta_y^{Zak} = \pi$) to form a super-cell, as demonstrated in Fig. \ref{Figure.2}(a). One unit cell of PC$_2$ is sandwiched between ten unit cells of PC$_3$, forming two coupled topological interfaces. The radius of scatterers along the coupled interfaces (i.e., scatterers in PC$_2$) is denoted as $r_\textrm{i}$. Compared to conventional topological interfaces, the coupled topological interfaces offer the superiority of multi-mode and broadband, which are beneficial for multi-mode rainbow trapping and broadband energy harvesting. The TISs in conventional topological interfaces are discussed in the Supplementary Material. Moreover, the coupled interface region provides sufficient space to host piezoelectric structures or materials for energy transduction. Fig. \ref{Figure.2}(b) illustrates the band structure of the PC super-cell. Black dots signify the bulk states, whereas red, yellow, and green dots highlight the TISs within the band gap. The pressure patterns of the three TISs at $k_x=\pi/a$ are plotted in Fig. \ref{Figure.2}(c). TISs 1, 2 and 3 exhibit S, S and A patterns with respect to the mirror plane (denoted by a black dashed line). To facilitate the experiment, TISs 1 and 2 are selected for realizing rainbow trapping.

\begin{figure*}[]
	\centering
	\includegraphics{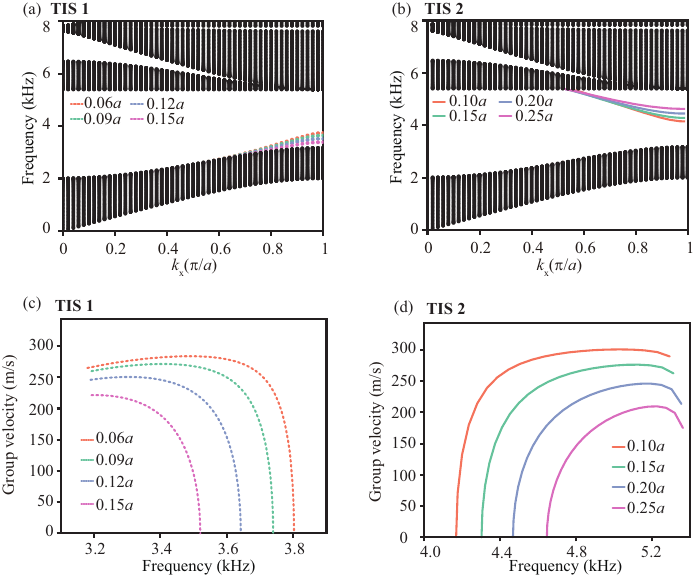}
	\caption{\label{Figure.3}  Dispersion curves of TISs 1 (a) and 2 (b) for different radii of scatterers along the coupled interfaces. Group velocity curves of TISs 1 (c) and 2(d) for different radii of scatterers along the coupled interfaces.}
\end{figure*}

Notably, in Figs. \ref{Figure.3}(a) and \ref{Figure.3}(b), the zero group velocity of TISs 1 and 2  appears at the Brillouin zone boundary $k_x$ = $\pi/a$. The radius of scatterers along the coupled interfaces $r_\textrm{i}$ are changed from $0.04a \sim 0.182a$, and $0.1a \sim 0.28a$ for TISs 1 and 2, respectively. As $r_\textrm{i}$ increase, the dispersion curve of TIS 1 (TIS 2) shifts to the lower (higher) frequency region. The group velocity  $v_\textrm{g}$ ($v_\textrm{g}$ = $\partial$$\omega$/$\partial$$k$, where $\omega$ is angular frequency and $k$ wave number) is affected by the geometry of the scatterers along the interfaces. The group velocity $v_\textrm{g}$ of TISs 1 and 2 as a function of $r_\textrm{i}$ is shown in Figs. \ref{Figure.3}(c) and \ref{Figure.3}(d), respectively. As $r_\textrm{i}$ increases, the cut-off frequencies (i.e., $v_\textrm{g}$$\rightarrow$ 0) of TISs 1 and 2 shift to lower and higher frequencies, respectively. Consequently, when acoustic waves enter into the PCs with zero group velocity, they are trapped and cannot propagate further.

\section{\label{Sec3}Topological rainbow trapping}

\begin{figure*}[]
	\centering
	\includegraphics{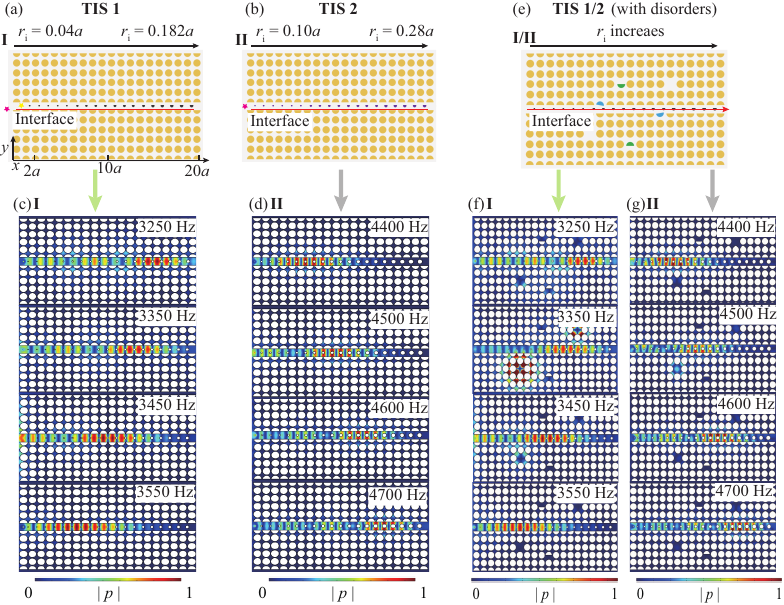}
	\caption{\label{Figure.4} Schematic diagrams of PC samples I (a), II (b), and with structural disorders (e). Topological rainbow trapping in PC samples I (c), II (d), and in perturbed PC samples I (f) and II (g) at different frequencies.}
\end{figure*}

In this section, two configurations of gradient PC structures (with a size of $20 \times 11 $ unit cells) are constructed to observe topological rainbow trapping. PC structures I and II correspond to TISs 1 and 2, respectively, as illustrated diagrammatically in Figs. \ref{Figure.4}(a) and \ref{Figure.4}(b). The radius of scatterers along the coupled interfaces $r_\textrm{i}$ increases linearly along the $+x$ direction. The expression of $r_\textrm{i}$ is described as $r_\textrm{i}$ = $r_0 + (n-1) \delta_r$, where $\delta_r$ is the step variation in $r_\textrm{i}$ and $n$ ($n = 1, 2, ..., 19$) denotes the position along the interfaces. The configuration of PC structure I (II) is given as follows: $r_0$ = $0.04 a$ and $\delta_r$ = $ 0.00789 a $ ($r_0$ = $ 0.1 a $ and $\delta_r$ = $ 0.01 a $). From Section \ref{Sec2}, it is known that if the radius of scatterers along the coupled interfaces $r_\textrm{i}$ decreases monotonously, the position of zero group velocity of TISs at different frequencies varies correspondingly. As a result, the gradient PC structures with coupled interfaces exhibit a fascinating phenomenon known as topological rainbow trapping. In the numerical simulations, a point monopole source  (marked by a star) is used to generate acoustic waves and placed on the left side of the PCs. Besides, perfectly matched layers (PMLs) are employed to reduce the acoustic wave reflection at the outer boundaries. To verify the robustness of the topological rainbow trapping, structural disorder is introduced into the scatterers of PC structures I and II in three different ways [see Fig. \ref{Figure.4}(e)]. The rotation angle $\varphi$ (marked in blue) and the shape (marked in green) are perturbed, as well as the absence of some scatterers is introduced, where $\varphi$ changes randomly from $-\ang{10}$ to $\ang{10}$. 

The frequency responses of the two PC structures with different gradient interfaces are calculated. First, for structure I, the acoustic pressure fields at four selected frequencies (3250, 3350, 3450 and 3550 Hz) are represented on Fig. \ref{Figure.4}(c). As expected, the concentration of acoustic energy along the coupled interfaces (marked by red lines) depends on the frequency of the incident waves. Acoustic waves are well-guided along the coupled interfaces, with waves at lower frequencies traveling further than those at higher frequencies. Due to the zero group velocity effect, acoustic waves at various frequencies localize at distinct spatial points, leading to the detection of the topological rainbow trapping effect arising from the gradient interfaces. Then, for structure II, the pressure profiles with incident waves at 4400, 4500, 4600 and 4700 Hz are presented in Fig. \ref{Figure.4}(d). Apparently, the topological rainbow trapping is also achieved in PC structure II. In comparison, the results for the perturbed PC structures are displayed in Figs. \ref{Figure.4}(f) and \ref{Figure.4}(g). Topological rainbow trapping demonstrates robustness against structural disorders, as the results with and without such disorders show a strong correlation.

\begin{figure*}[]
	\centering
	\includegraphics{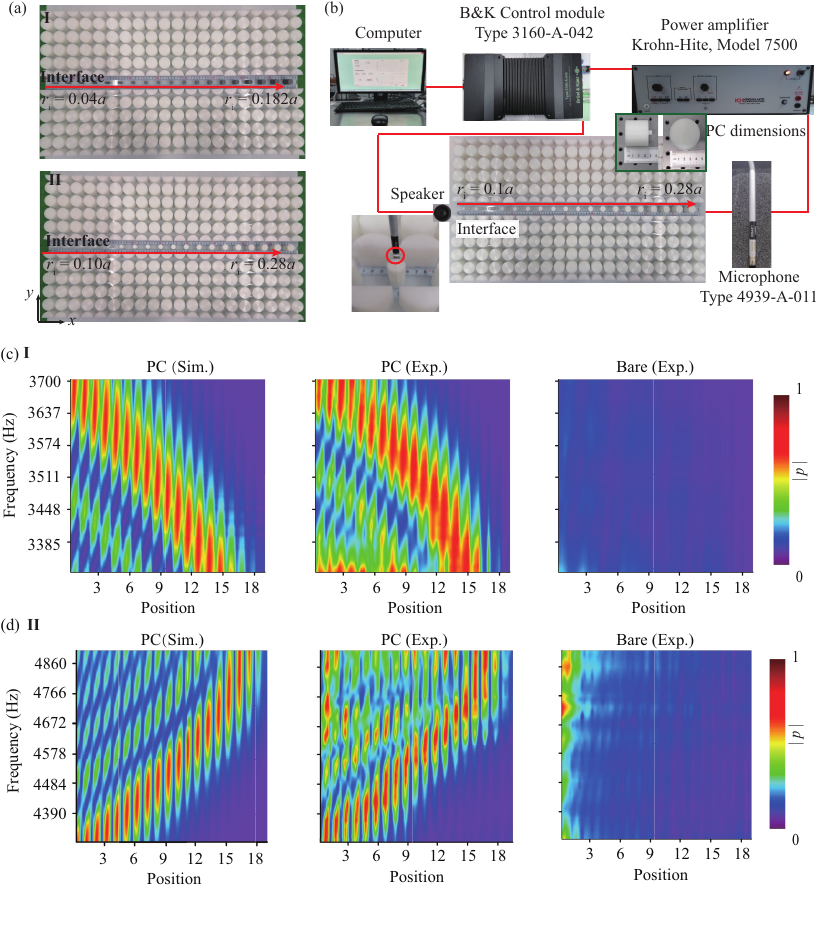}
	\caption{\label{Figure.5}(a) Photographs of the 3D-printed PC samples (I and II). (b) Photograph of the experimental setup and the inset reveals the enlarged view of the drilled holes for inserting the microphone (red wireframe) and the PC dimensions (green wireframe). Pressure profiles within the flawless PC samples and the bare structure are obtained through numerical simulations and experimental measurements: (c) sample I and (d) sample II.}
\end{figure*}

For experimental confirmation, polymer PC samples are produced through 3D printing techniques. Photographs of these manufactured PC samples are given in Fig. \ref{Figure.5}(a). The gradient PC samples consist of $20 \times 11$ unit cells and the radius of scatterers on the coupled interfaces $r_\textrm{i}$ varies progressively along the $+x$ axis in agreement with the simulations [see Figs. \ref{Figure.4}(a) and \ref{Figure.4}(b)]. The experimental arrangement is shown in Fig. \ref{Figure.5}(b). The PC samples are positioned in a 35 mm high planar waveguide to meet the requirements for a 2D approximation, with the top and bottom sandwiched between PMMA and PE plates, respectively. To accurately measure the acoustic field, 19 holes are made in the top  PMMA layer of the planar waveguide to hold the microphones. These holes allow for easy insertion of the microphones, enabling efficient detection of the sound pressure along the PC interfaces. A signal generator module produces the signals within the specified frequency range, which are then amplified by a power amplifier. These signals are  emitted through a loudspeaker positioned to the left of the fabricated PC interfaces. Additionally, a bare structure (planer waveguide without PCs) is also considered for comparison.

Figures \ref{Figure.5}(c) and \ref{Figure.5}(d) indicate the numerical and experimental findings on topological rainbow trapping in PC structures I and II, delineated by frequency. Herein, we will discuss the results of PC structure I in detail. Obviously, the acoustic waves at lower frequencies (e.g., 3385 Hz) can pass through the entire interface of the graded PC structure. In contrast, the acoustic waves at higher frequencies (e.g., 3700 Hz) can only travel a short distance before being stopped owing to the rainbow trapping effect. The measured and calculated results match well, although the location of the high pressure region in the experiment changes slightly from the numerical prediction. On the other hand, the rainbow trapping of PC structure II is also verified experimentally, as displayed in Fig. \ref{Figure.5}(d). Importantly, compared with the sound pressure in the bare structure, the topological rainbow trapping in the designed PC structures has a significant amplification  [see Figs. \ref{Figure.5}(c) and \ref{Figure.5}(d)].

\begin{figure*}[]
	\centering
	\includegraphics{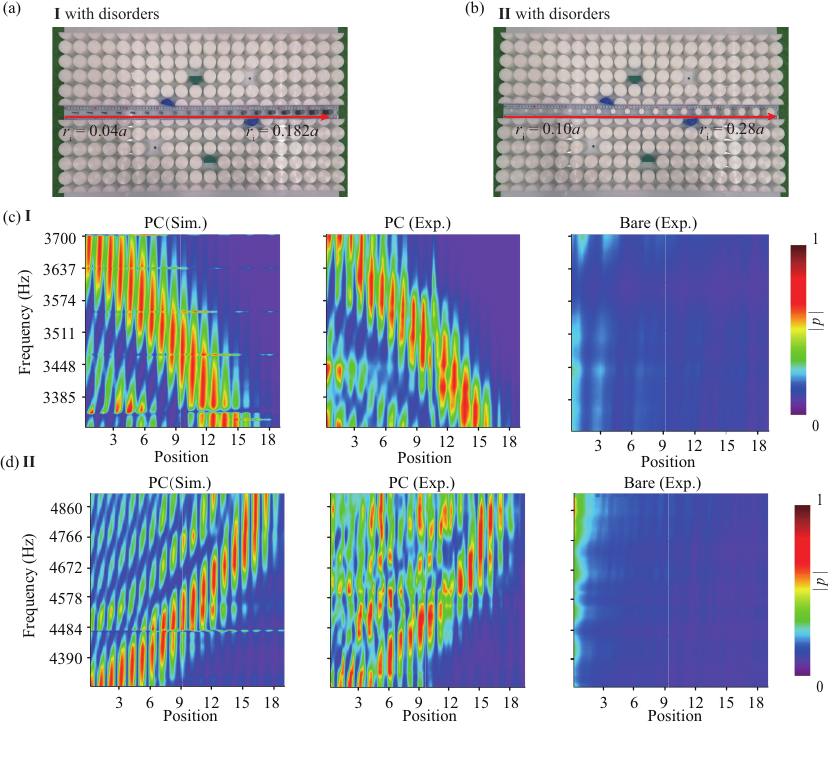}
	\caption{\label{Figure.6} Photographs of PC samples I (a) and II (b) with several structural disorders, where the scatterers are randomly modulated in rotation angle (highlighted by blue color), shape (highlighted by green color) and absence. Numerical and experimental pressure distributions in perfect PC  and the bare structure: (c) sample I and (d) sample II.}
\end{figure*}

Furthermore, to demonstrate the robustness of topological rainbow trapping and acoustic energy amplification, several structural disorders are introduced into PC structures I and II, as shown in Figs. \ref{Figure.6}(a) and \ref{Figure.6}(b). The introduced disorders align with the numerical simulations in Fig. \ref{Figure.4}(c). Figures \ref{Figure.6}(c) and \ref{Figure.6}(d) display the simulated and observed pressure distributions in the corresponding frequency ranges for the two disordered PC structures. The numerically calculated and experimentally measured results for topological rainbow trapping are in good agreement. There is a slight deviation at the high pressure position resulting from experimental discrepancies. Comparing Figs. \ref{Figure.5} and \ref{Figure.6}, we infer that the rainbow trapping of the TISs remains despite the introduction of different types of structural disorders.  

\section{\label{Sec4}Experimental validation of acoustic energy harvesting}

\begin{figure*}[]
	\centering
	\includegraphics{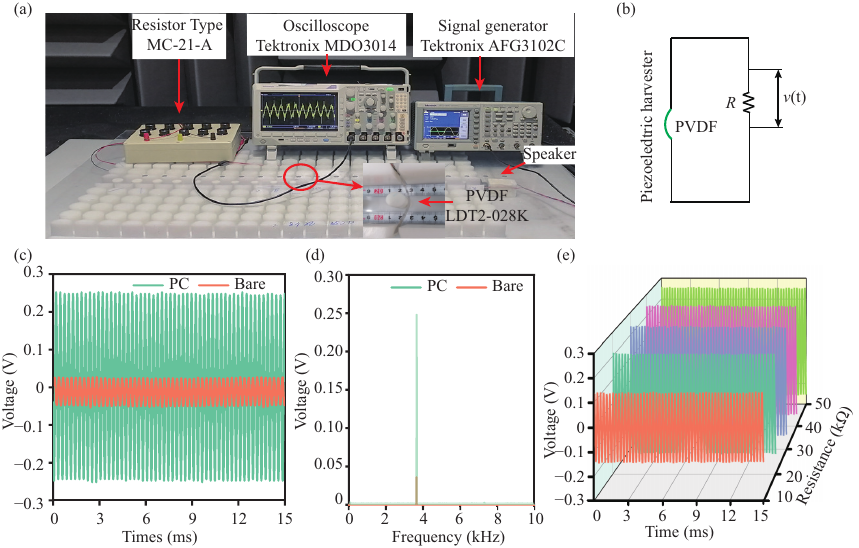}
	\caption{\label{Figure.7} (a) Illustration of the experimental arrangement used for energy harvesting. (b) Schematic representation of the piezoelectric energy harvesting system. Voltage outputs recorded in both time (c) and frequency (d) domains. (e) Time-domain voltage outputs for various resistance loads.}
\end{figure*}

Displayed in  Fig. \ref{Figure.7}(a) is the setup for harvesting piezoelectric energy from acoustic waves, utilizing the topological rainbow trapping effect within gradient PCs. A layer of PVDF film is strategically positioned at the interfaces of PC structures to facilitate the transformation of acoustic energy into electrical energy [see the inset in Fig. \ref{Figure.7}(a)]. A schematic of the piezoelectric energy harvesting system circuit is depicted in Fig. \ref{Figure.7}(b). An acoustic signal is generated by a signal generator and subsequently amplified by a power amplifier. Voltage measurements across the resistor are captured via an oscilloscope. To ascertain the effectiveness of the PC energy harvester, a continuous sine-wave signal is employed, with the frequencies corresponding to the topological rainbow trapping.

Voltage measurements from the piezoelectric film in topological PC structure I at 3675 Hz over time are displayed in Fig. \ref{Figure.7}(c). The electric output of the bare structure (without PCs) is measured as the reference. Additionally, Fig. \ref{Figure.7}(d) displays the corresponding frequency-domain voltage results. Clearly, the output voltage of the topological PC energy harvester is superior to that of the bare device, due to the enhanced acoustic pressure from the topological rainbow trapping effect.  Notably, the load resistance greatly affects the output of the phononic crystal energy harvester. Fig. \ref{Figure.7}(e) illustrates the voltage outputs measured from the PC energy harvester under various external load resistances. Clearly, the voltage amplitude increases with the load resistance.

\begin{figure*}[]
	\centering
	\includegraphics{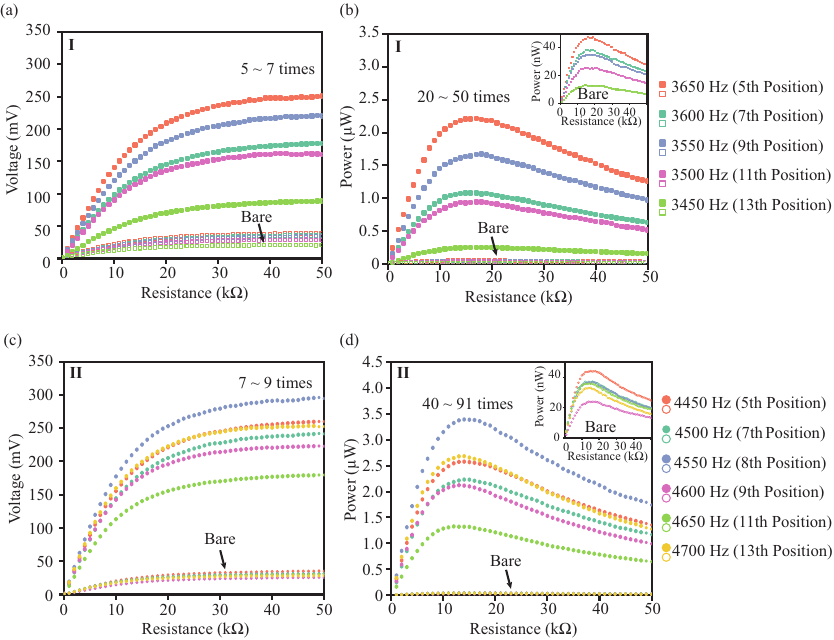}
	\caption{\label{Figure.8}Measured voltage (a, b) and power (c, d) outputs relative to load resistance at selected frequencies in  PC structures I and II. Dots represent the outputs from the PC structure, while circles represent those from the bare structure. The results of the bare structure are visually enlarged in the insets of (b) and (d).}
\end{figure*}

To verify that the proposed PC structures based on the topological rainbow trapping can achieve broadband energy harvesting of acoustic waves, the piezoelectric films are placed at different positions along the structural interfaces, as shown in  Fig. \ref{Figure.7}(a). Since the wave characteristics of the topological rainbow trapping in both the spatial and frequency domains, as well as its robustness against structural disorders, have been discussed in Section \ref{Sec3} (see Figs. \ref{Figure.5} and \ref{Figure.6}), this section focuses on efficacy of acoustic energy harvesting. Five frequencies (for PC structure I) from 3450 to 3650 Hz and six frequencies (for PC structure II) ranging from 4450 to 4700 Hz are considered in the energy harvesting experiments. Accordingly, only five positions for structure I (six positions for structure II) along the interfaces are employed to host the piezoelectric films. Notably, due to the restriction on the position of piezoelectric films, the excited frequencies are generally not the exact ones needed for the optimal energy harvester. Figs. \ref{Figure.8}(a) and \ref{Figure.8}(c) depict the voltage output from the PC and reference energy harvesters relative to varying load resistances. As resistance escalates, the voltage rises steadily, though the acceleration of increase slows and eventually plateaus at higher resistances. Notably, the output voltage from the topological PC structures multiplies by approximately 5 to 7 times in structure I (i.e., TIS 1) and 7 to 9 times in structure II (i.e., TIS 2), effectively harnessing acoustic energy across all frequencies evaluated. The calculation of power from the piezoelectric harvester follows Ohm’s law, represented as $P$ = $U^2$/$R$, with $P$ indicating power, $U$ voltage, and $R$ the load resistance. The power output increases with rising load resistance, peaks at a certain point, and then begins to decrease, as presented in Figs. \ref{Figure.8}(b) and \ref{Figure.8}(d). The gradient PC structures produce up to 3.5 $\mu$W of power at 4550 Hz with a 14 k$\Omega$ external load [see Fig. \ref{Figure.8}(d)]. Herein, the amplification ratio is the power output comparison between the PC and the bare reference structures. Remarkably, energy harvesting is achieved from 3300 to 3800 Hz (from 4350 to 4900 Hz) in topological PC structure I (II). The amplification ratios are $20\sim50$ and $40\sim91$ for the proposed PC harvesters I and II, respectively.

\section{\label{Sec5}Conclusion}

In this work, we introduce PC structures with gradient coupled interfaces designed to facilitate topological rainbow trapping and enhance acoustic energy harvesting. The proposed 2D PCs are constructed using polymeric scatterers in the air background, displaying a broad band gap characterized by different Zak phases. Thanks to the geometric symmetry, the Zak phase can be tuned by varying the PC unit cells. TISs develop where two PCs with differing Zak phases meet. The results reveal that acoustic waves at different frequencies diverge, stop, and are enhanced at distinct locations within the coupled interfaces. In addition, the topological rainbow trapping is robust to randomly structural disorders. Subsequently, A PVDF film is integrated into the coupled interfaces of gradient PCs, enabling efficient acoustic-to-electrical energy conversion. The proposed PC structure exhibits the capacity of broadband energy harvesting. The maximum output power of the topological PC energy harvester is 91 times greater than that of the bare structure. Notably, the design of coupled interfaces possess the characteristics of multi-mode and broad bandwidth, providing wide application prospects for energy harvesting, sensing and selective filtering.

\section*{CRediT authorship contribution statement}

Xiao-Lei Tang: Data curation, Formal analysis, Investigation, Methodology, Validation, Visualization, Writing - original draft;  Xue-Qian Zhang: Data curation, Investigation; Tian-Xue Ma: Conceptualization, Supervision, Formal analysis, Validation, Writing - review \& editing; Miso Kim: Funding acquisition, Resources, Writing - review \& editing; Yue-Sheng Wang: Funding acquisition, Supervision, Resources, Writing - review \& editing.

\section*{Declaration of Competing Interest}
The authors declare that they have no known competing financial interests or personal relationships that could have appeared to influence the work reported in this paper.

\section*{Acknowledgments}
 We sincerely appreciate Innovative Research Group of NSFC (Grant No. 12021002), National Natural Science Foundation of China (Grant No. 12372087), and the National Research Foundation of Korea grant funded by the Korea government (No.RS-2023-00254689) for the financial support to this study. Xiaolei-Tang is grateful to the support of the China Scholarship Council (Grant No. 202306250172).

\section*{DATA AVAILABILITY}
The data that support the findings of this study are available from the corresponding author upon reasonable request.

\bibliographystyle{elsarticle-num}
\biboptions{numbers,sort&compress}
\bibliography{Manuscript}


\end{document}